\documentclass[8.5pt,twoside,twocolumn]{article}
\usepackage[super,sort&compress,comma]{natbib}
\usepackage{widetext}

\usepackage[version=3]{mhchem}
\usepackage{times,mathptmx}
\usepackage{sectsty}
\usepackage{balance}

\usepackage{graphicx} 
\usepackage{lastpage}
\usepackage[format=plain,justification=raggedright,singlelinecheck=false,font=small,labelfont=bf,labelsep=space]{caption}
\usepackage{fancyhdr}
\usepackage{amssymb,amsmath}
\begin{document}

\thispagestyle{plain}
\fancypagestyle{plain}{
\fancyhead[L]{\includegraphics[height=8pt]{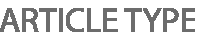}}
\fancyhead[C]{\hspace{-1cm}\includegraphics[height=20pt]{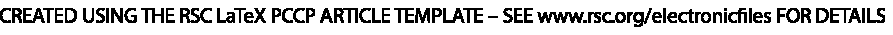}}
\fancyhead[R]{\includegraphics[height=10pt]{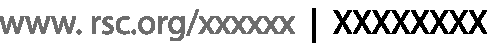}\vspace{-0.2cm}}
\renewcommand{\headrulewidth}{1pt}}
\renewcommand{\thefootnote}{\fnsymbol{footnote}}
\renewcommand\footnoterule{\vspace*{1pt}%
\hrule width 3.4in height 0.4pt \vspace*{5pt}}

\setcounter{secnumdepth}{5}

\makeatletter
\def\subsubsection{\@startsection{subsubsection}{3}{10pt}{-1.25ex plus -1ex minus -.1ex}{0ex plus 0ex}{\normalsize\bf}}
\def\paragraph{\@startsection{paragraph}{4}{10pt}{-1.25ex plus -1ex minus -.1ex}{0ex plus 0ex}{\normalsize\textit}}
\renewcommand\@biblabel[1]{#1}
\renewcommand\@makefntext[1]%
{\noindent\makebox[0pt][r]{\@thefnmark\,}#1}
\makeatother
\renewcommand{\figurename}{\small{Fig.}~}
\sectionfont{\large}
\subsectionfont{\normalsize}

\fancyfoot{}
\fancyfoot[LO,RE]{\vspace{-7pt}\includegraphics[height=9pt]{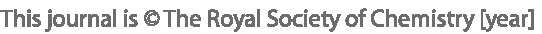}}
\fancyfoot[CO]{\vspace{-7.2pt}\hspace{12.2cm}\includegraphics{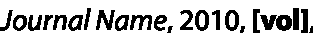}}
\fancyfoot[CE]{\vspace{-7.5pt}\hspace{-13.5cm}\includegraphics{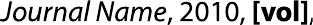}}
\fancyfoot[RO]{\footnotesize{\sffamily{1--\pageref{LastPage} ~\textbar  \hspace{2pt}\thepage}}}
\fancyfoot[LE]{\footnotesize{\sffamily{\thepage~\textbar\hspace{3.45cm} 1--\pageref{LastPage}}}}
\fancyhead{}
\renewcommand{\headrulewidth}{1pt}
\renewcommand{\footrulewidth}{1pt}
\setlength{\arrayrulewidth}{1pt}
\setlength{\columnsep}{6.5mm}
\setlength\bibsep{1pt}

\twocolumn[
 \begin{@twocolumnfalse}
\noindent\LARGE{\textbf{Tunable Slow Dynamics in a New Class of Soft Colloids}}
\vspace{0.6cm}

\noindent\large{\textbf{Federica Lo Verso,\textit{$^{a}$} Jos\'e A. Pomposo,\textit{$^{a,b,c}$} Juan Colmenero,\textit{$^{a,b,d}$} and Angel J. Moreno$^{\ast}$\textit{$^{a,d}$}}}\vspace{0.5cm}

\noindent \normalsize{                                                                                                               
By means of extensive simulations, we investigate concentrated
solutions of globular single-chain nanoparticles (SCNPs), an emergent class of synthetic soft nano-objects. 
By increasing the concentration, the SCNPs show a reentrant behaviour in their structural
and dynamical correlations, as well as a soft caging regime and weak dynamic heterogeneity.
The latter is confirmed by validation of the Stokes-Einstein relation up to concentrations far beyond the overlap density. 
Therefore SCNPs arise as a new class of soft colloids, exhibiting slow dynamics and 
actualizing in a real system  structural and dynamical anomalies proposed by models of ultrasoft particles. 
Quantitative differences in the dynamical behaviour depend
on the SCNP deformability, which can be tuned through the degree of internal cross-linking. }
\vspace{0.5cm}
 \end{@twocolumnfalse}
  ]

\footnotetext{\textit{$^{a}$~Centro de F\'isica de Materiales (CSIC, UPV/EHU) and Materials Physics Center MPC, 
Paseo Manuel de Lardizabal 5, E-20018 San Sebasti\'{a}n, Spain.}}
\footnotetext{\textit{$^{b}$~Departamento de F\'isica de Materiales, Universidad del Pa\'is Vasco (UPV/EHU), Apartado
1072, E-20080 San Sebasti\'an, Spain.}}
\footnotetext{\textit{$^{c}$~IKERBASQUE - Basque Foundation for Science, Mar\'ia D\'iaz de Haro 3, E-48013 Bilbao,
Spain.}}
\footnotetext{\textit{$^{d}$~Donostia International Physics Center (DIPC), Paseo Manuel de Lardizabal 4, E-20018
San Sebasti\'{a}n, Spain. }}
\footnotetext{\textit{$^{\ast}$~Tel: +34 943018845; E-mail: $angeljose.moreno@ehu.es$}}

\section{Introduction}
												                     
Precisely defined polymers, folded into functional nanostructures that are able to target complex tasks, are constantly encountered in nature. 
In order to engineering functional soft nanomaterials that closely mimic biomolecules in structure and behaviour, a paradigm in polymer synthesis involves handling single polymer chains \cite{cite-key}. 
Among the various techniques employed to these ends, the collapse of single polymer chains (precursors)
via purely intramolecular cross-linking, into single-chain nanoparticles (SCNPs),  has 
gained increasing interest over recent years  \cite{altintas2012single,C5CS00209E,doi:10.1021/acs.chemrev.5b00290,doi:10.1021/acs.macromol.5b01456}.
Significant effort is being devoted to endow SCNPs with useful functions for multiple
applications in, e.g., nanomedicine, biosensing, bioimaging or catalysis 
\cite{terashima2011,gillissen2012single,perez2013endowing,sanchez2013design,MARC:MARC201500547}.

Recent investigations by experiments and simulations have highlighted that, in general, the usual synthesis protocols in good solvent conditions
produce SCNPs with open, sparse conformations \cite{perez2013endowing,moreno2013,sanchez2013design,pomposo2014ACS}.
As revealed by extensive simulations \cite{moreno2013}, this feature originates from a fundamental property: the self-avoiding conformations
of the linear precursors in 
good solvent. In these conditions the formation of long loops, at the origin of an efficient global compaction of the SCNP, is statistically unfrequent.
Instead, bonding between reactive groups at short contour distances, which promotes only     
local globulation, is highly favoured.

The control of the size, shape and internal malleability of SCNPs is a key issue, since their target functions
will intimately depend on such parameters.
To bypass the intrinsic limitations of the synthesis in good solvent, and to obtain globular SCNPs, we have recently designed solvent-assisted synthesis routes \cite{C4SM02475C} that promote bonding over long contour distances. 
These experimentally accessible protocols \cite{doi:10.1021/ja00897a025,MARC:MARC201600139}, 
described in detail in Ref.~\cite{C4SM02475C}, prevent interparticle aggregation and do not require specific sequence control of the precursors. In all cases, after completing the cross-linking and restoring good solvent conditions, the swollen SCNPs are globular objects. It is worth noting that, even by starting from the same precursors (same molecular weight and fraction of reactive groups), the topologies of the obtained SCNPs are intrinsically polydisperse. 
Still, the former routes lead to much narrower distributions in size and shape than those obtained 
by performing the synthesis in good solvent \cite{C4SM02475C}. Recently, we have introduced a new route in {\it good} solvent, 
where formation of long loops is promoted by using long multifunctional cross-linkers. 
Simulations and experiments confirm that this route 
also produces globular SCNPs \cite{doi:10.1021/ma5017133}.

All in all, the  progress  made so far in the synthesis and characterisation  disclose SCNPs  as  versatile systems  
allowing for effective tuning of their softness and folding, and open up new realms for understanding and tailoring 
the phase and dynamical behaviour of complex fluids.
In this article we focus on the case of
globular SCNPs in concentrated solutions.
By means of large-scale simulations
based on a standard bead-spring model \cite{kremer1990dynamics}, 
we investigate the structure  and dynamics of these solutions.
In analogy with models of ultrasoft particles \cite{berthier-hertz},
by increasing the concentration the SCNPs show a reentrant behaviour in their structural
and dynamical correlations, as well as a soft caging regime and weak dynamic heterogeneity.
These anomalies quantitatively depend on the SCNP degree of deformability. 
Globular SCNPs, with a tunable degree of compactness, emerge as a new class of soft colloids
bridging the gap between linear polymers and hard colloids.
As a consequence they are a very promising, experimentally realisable system
for getting insight into the mechanisms of diffusion of soft nano-objects in crowded environments
(as expected for nanocarriers in biological habitat) \cite{blanco,zhang},
and to draw new strategies for tailoring the rheological properties
of polymer-based nanomaterials \cite{Vlassopoulos2014561}.

The article is organized as follows. In Section 2 we give model and simulation details. In Sections 3 and 4 we present
and discuss structural and dynamic anomalies of concentrated solutions of SCNPs. Conclusions are given in Section 5.

\section{Model and Simulation Details}

We use a bead-spring model \cite{kremer1990dynamics,C4SM02475C} to simulate the SCNPs. 
We consider two different types of SCNPs, denoted as P30 and P72. Each SCNP of type P30 consists of 520 beads in total. $N = 400$ beads 
correspond to the backbone chain and $L = 120$ beads are reactive side groups, so that the fraction of reactive side groups is $f = L/N =$ 30 \%. 
Each SCNP of type P72 is made of 688 beads in total, with $N = 200$ beads forming the backbone and a fraction $f =$ 72 \% of reactive side groups, with the reactive beads at the ends of the groups.  
In all cases the side groups are randomly distributed along the backbone contour of the precursor. 

The SCNPs are synthesised from the precursors by following the protocols introduced in Ref.~\cite{C4SM02475C}, which are designed to prevent intermolecular aggregation,
i.e., cross-linking is purely intramolecular by construction. Briefly, in the synthesis of the P72-SCNPs, cross-linking is performed in bad solvent, with the precursors anchored to a surface. 
In the synthesis of the P30-SCNPs  we use an amphiphilic precursor with unreactive solvophilic and reactive solvophobic units, which forms a single core-shell structure with the reactive units in the core.
Cross-linking is irreversible in all cases, i.e., when two reactive groups form a mutual bond they remain permanently bonded and are not allowed to
form new bonds with other groups. 
After completing the cross-linking and restoring good solvent conditions for all the units, the swollen SCNPs obtained by both routes are globular objects.
Further details of the model and cross-linking procedure can be found in Ref.~\cite{C4SM02475C}. 

We use each type of SCNP (P30, P72) to construct their corresponding solutions, 
which are investigated at different concentrations.
In these solutions the interactions between the beads are modeled by a purely repulsive Lennard-Jones (LJ) potential:
\begin{equation}
V_{\rm LJ}(r)= 4\epsilon [(\sigma/r)^{12} -(\sigma/r)^6 +1/4] ,
\end{equation}
which is cut-off at $r = 2^{1/6}\sigma$.
We use a mass $m=1$ for all beads, and set the LJ parameters \cite{kremer1990dynamics,C4SM02475C} $\epsilon = \sigma = 1$ as units of energy and distance respectively. 
Interactions between mutually bonded beads are modeled by a FENE potential \cite{kremer1990dynamics}:
\begin{equation}
V_{\rm FENE}(r)= -\epsilon KR_0^2 \ln[ 1-(r/R_0\sigma)^2 ] ,
\end{equation}
with $K =15$ and $R_0 = 1.5$.
The use of the LJ and FENE potentials prevents strong fluctuations of the bonds and guarantees
chain uncrossability \cite{kremer1990dynamics}. With this, the interactions between the SCNPs in the solutions are driven just by monomer 
excluded volume and bond uncrossability, mimicking good solvent conditions. 

Since, as aforementioned, cross-linking of the precursors leads to SCNPs with topological polydispersity,
the corresponding solutions P72 and P30 are intrinsically polydisperse.   
We investigate a third solution (M72) formed by replicas of the {\it same} SCNP. The replicated SCNP is taken from the P72 system. 
Namely we select a SCNP with
the same time-averaged size and shape as the corresponding averages
over all the SCNPs (see below).
The system M72 is intrinsically monodisperse: all replicas are topologically identical, and polydispersity just originates from the intramolecular dynamics that leads to size and shape fluctuations of the different replicas. By investigating the P72 and M72 systems, we discriminate the role of the intramolecular fluctuations from that of the inherent topological polydispersity.
The P72-SCNPs are more tightly cross-linked, and hence less deformable, than the P30-SCNPs. 
Therefore, by comparing them we investigate the role of the intramolecular deformability on the structure and dynamics of the solution.

In the three investigated solutions the used number of SCNPs is  $N_{\rm p} = 125$, yielding a total number of monomers $N_{\rm mon} = 86000$ (P72 and M72) and  $N_{\rm mon} = 65000$ (P30).
The SCNPs are initially inserted in a cubic box of size $L$, with periodic boundary conditions and monomer density $\rho_{\rm m} = N_{\rm mon}/L^3  \sim  0.015$. This density is about 20 times smaller than the overlap concentration (see below) and is considered as `infinite dilution'  $\rho \rightarrow 0$. The minimum distance between any two SCNPs inserted in the initial configuration is chosen in order to prevent concatenations. Once the box at $\rho \rightarrow 0$ is constructed and equilibrated, it is very slowly compressed (again preventing concatenation) and equilibrated at different selected densities. Equilibration runs extend over sufficiently long times so that each SCNP diffuses to a distance of several times its diameter, fully decorrelating from its initial conformation. The equilibrated boxes are used for the acquisition runs to characterize the conformational and dynamical properties of the SCNPs. 

\begin{figure}
\includegraphics[width=0.91\linewidth]{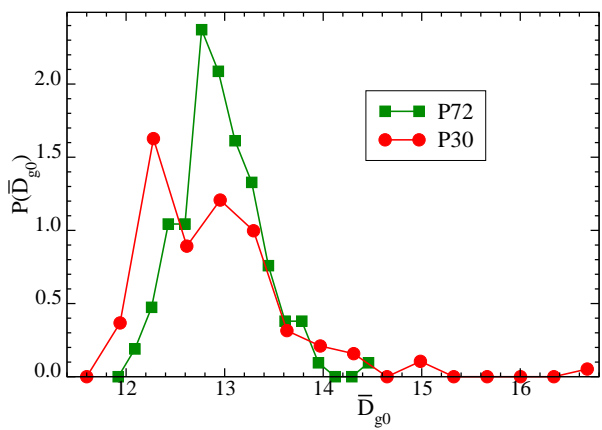}			    
\caption{Distribution of time-averaged diameters of gyration, $\bar{D}_{\rm g0}$, for the systems P72 and P30 at infinite dilution}	
\label{distdg}		
\end{figure}												          

The SCNP density, normalized by the average diameter of gyration at infinite dilution ($D_{\rm g0}$), 
is defined as $\rho = N_{\rm p}D_{\rm g0}^3/L^3$. For all the investigated systems $D_{\rm g0} \approx 12.8$ 
(see mean value of the distribution
of time-averaged diameters of gyration, Fig.~\ref{distdg}). The overlap density is defined as $\rho = \rho^{\ast} = 1$ 
and corresponds to the density for the onset of intermolecular contacts at the macromolecular peripheria. The investigated densities cover the whole range from infinite dilution to values of $\rho  = 1.6 - 2.7$ (depending on the specific system) well beyond the overlap density. 

The simulations have been peformed under Langevin dynamics (LD)
(see Ref.~\cite{moreno2013} for details of the implementation) at temperature $T = 1$. 
Equilibration and acquisition runs typically extend over $10^8$ time steps. The simulations of the solutions have been performed by using 
the GROMACS 4.6.5 package \cite{BERENDSEN199543}. To improve statistics of the dynamic observables 
(mean squared displacements, diffusivities and scattering functions), additional averages have been performed over 10 time origins. 
No signatures of aging (i.e., drift of properties with the time origin) have been observed.

The use of LD neglects hydrodynamic effects, whose implementation involves a much higher computational cost. 
In order to have a cogent test of this assumption, we have performed additional simulations with hydrodynamic interactions for the system P30 at the lowest density, and compared them with the simulations under LD at the same density. Indeed P30 represents the less compact and more flexible system considered in this work, and consequently the one which should be more affected by hydrodynamic interactions, which moreover  become more relevant at lower concentrations. To simulate hydrodynamic interactions we implement the multi-particle collision dynamics (MPCD) scheme, which leads to correct hydrodynamics (Navier-Stokes equation) 
in the continuum limit \cite{doi:10.1021/jp046040x}. Details on the implementation of MPCD can be found in, 
e.g., Refs.~\cite{mussawisade-hydro,sing-hydro}. The solvent is represented as $N_{\rm s}$ point particles of mass $m_{\rm s}$. The MPCD algorithm consists of a streaming and a collision step. In the streaming step the solvent particles perform ballistic motion. In the collision step the solvent particles and the beads of the SCNPs are sorted into cells of size $d$. The corresponding velocities are rotated an angle $\alpha$, around a random axis, respect to the center-of-mass velocity of the cell. We use a density of solvent particles $N_{\rm s}/L^3 = 10$, and the parameters $m_{\rm s} = 0.1$, $\alpha = 150^{\circ}$  and $d = 1.125$ for the MPCD algorithm. The SCNPs in the MPCD simulations are distributed over independent boxes of size $L \sim 5D_{\rm g0}$, which prevents significant size effects on the hydrodynamic properties \cite{sing-hydro}.  Collisions as described above are performed every 50 MD steps. The monomers are propagated under LD between consecutive collision events, with a time step $\Delta t = 0.003$. With the former conditions, the collisional viscosity dominates over the kinetic viscosity, 
and hydrodynamic interactions are fully developed \cite{mussawisade-hydro,sing-hydro}.

\section{Structure}

\begin{figure}
\includegraphics[width=0.92\linewidth]{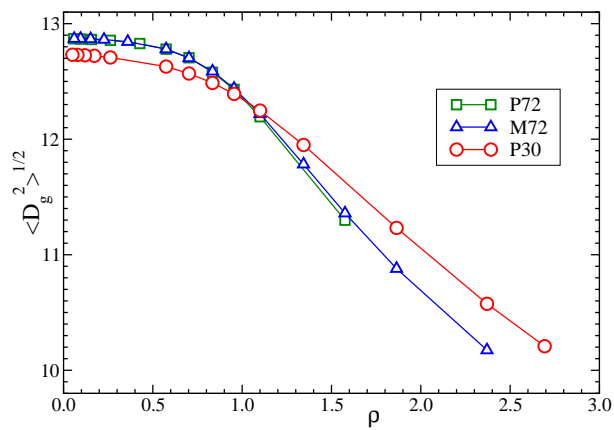}			    
\caption{Average diameter of gyration versus density for the systems P72, M72 and P30.}	
\label{dg-dens}		
\end{figure}												          

\begin{figure}
\includegraphics[width=0.96\linewidth]{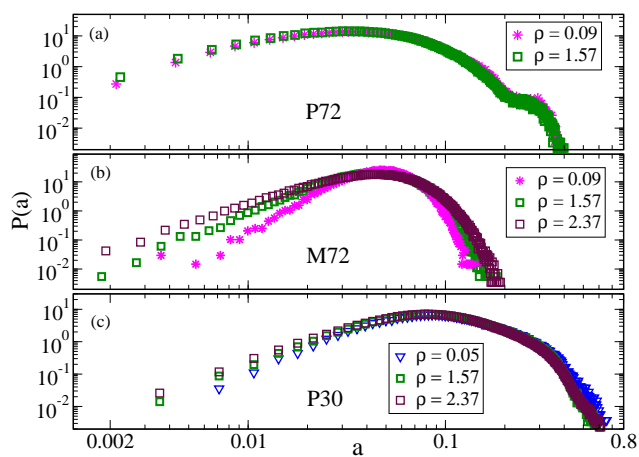}			    
\caption{Distributions of the asphericity for the three systems investigated, P72 (a), M72 (b), and P30 (c), for some representative densities. Symbol codes have the same meaning in all panels (see legends).}	
\label{dist-asph}		
\end{figure}

\begin{figure}
\includegraphics[width=0.96\linewidth]{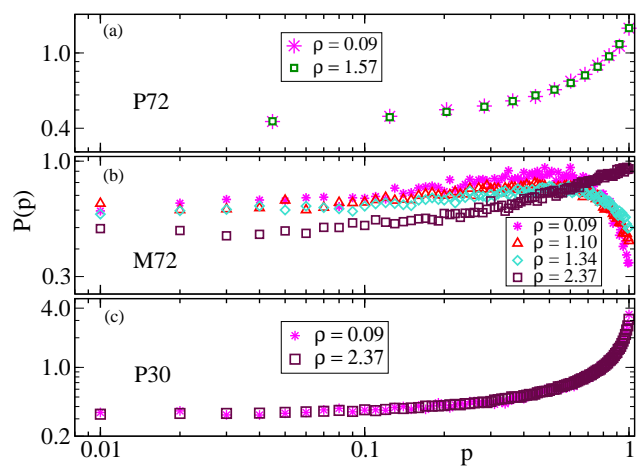}			    
\caption{Distributions of the prolateness for the three systems investigated, P72 (a), M72 (b), and P30 (c), for some representative densities. Symbol codes have the same meaning in all panels (see legends).}	
\label{dist-prol}		
\end{figure}

To highlight the soft colloidal character of the globular SCNPs, we investigate the density dependence
of the structural and dynamical correlations
of their centers-of-mass, from high dilution to concentrations beyond the overlap density.
Figs.~\ref{dg-dens},~\ref{dist-asph}, and \ref{dist-prol}
show the effect of density on the size and shape of the SCNPs. As expected, the SCNPs shrink at concentrations 
beyond the overlap density (Fig.~\ref{dg-dens}). Their size decreases up to about 25 \% at the highest investigated densities.

The fluctuations in the shape of the SCNPs can be characterized by the instantaneous values
of their asphericity ($a$) and prolateness ($p$) parameters \cite{moreno2013,C4SM02475C}.
The corresponding distributions are displayed in Figs.~\ref{dist-asph} and \ref{dist-prol}, respectively.
Results are shown for different densities, including infinite dilution ($\rho \ll \rho^{\ast}$) 
and the highest investigated density in each case.
We observe that SCNPs maintain their predominantly prolate ($p \rightarrow 1$) 
and quasi-spherical ($a \rightarrow 0$) character. Unlike for the size, the distributions of the shape parameters
are esentially unaffected by increasing the concentration in the polydisperse systems P72 and P30.
Visible, though moderate, changes are instead found in the monodisperse solution M72. Increasing the concentration far beyond the overlap density (see data for $\rho = 2.37$) results in a broader distribution of the asphericity and, on average, more prolate SCNPs. 
Apparently, if SCNPs are surrounded by other topologically different SCNPs, they can mutually optimize packing through internal deformations without altering their shape distribution. This is not possible if all SCNPs are topologically identical and have the same deformability, 
so that they optimize the packing by adopting, on average, more elongated conformations.


\begin{figure}
\includegraphics[width=0.93\linewidth]{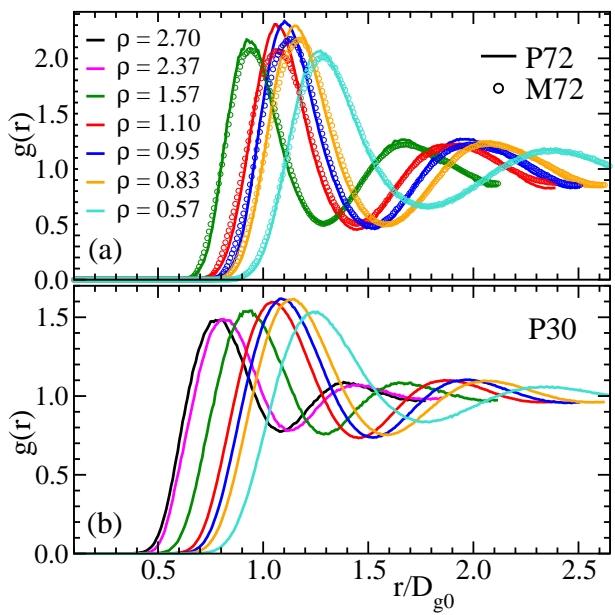}
\caption{Lines: radial distribution function of the centers-of-mass at different densities for the P72 (a) and P30 (b) solutions.
Circles in (a) are the results for M72. Same color codes in both panels.}
\label{gr}
\end{figure}

Fig.~\ref{gr} shows the radial distribution function, $g(r)$, of the SCNP centers-of-mass. 
In panel (a) we compare results for the polydisperse P72 solution and for its monodisperse counterpart M72.
Panel (b) shows results for P30.    
As expected, below the overlap concentration 
the height of the peaks grow with density, indicating increasing order.
However, at high concentration, we find two important features.
First, the system  always  remains  fluid. It does not crystallize even in the monodisperse case M72. 
This becomes evident by inspection of the mean-squared displacement (MSD), 
which does not saturate to a plateau and ultimately reaches the diffusive regime (Fig.~\ref{msd}a).
Second, above $\rho^{\ast}$ the first peak of $g(r)$
decreases, highlighting a loss of structural order 
in the solution by {\it increasing} the density beyond the overlap concentration.
This behaviour resembles that observed in some real soft colloids and models of ultrasoft particles
\cite{PhysRevLett.80.4450,doi:10.1021/jp808203d,PhysRevLett.90.238301,pamies-hertz,jacquin-harmon,berthier-hertz}. 
It originates from the penetrable character of the SCNPs that is inherent to their deformability. 
Thus, the observed loss of order at high density involves a broader distribution of interparticle 
distances and therefore strong interpenetration for the closest SCNPs. Unlike in  hard colloids, the
energetic cost for interprenetration is moderate, and is compensated by the entropic gain produced by disordering.

\begin{figure}
\includegraphics[width=.99\linewidth]{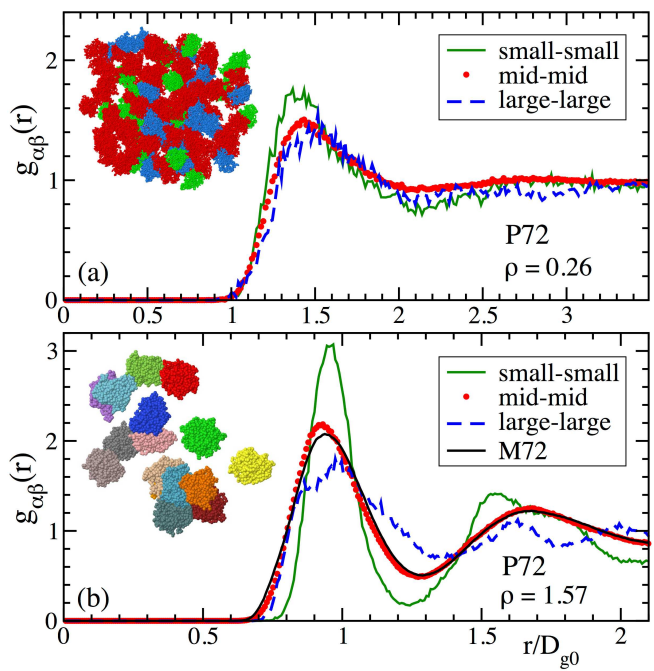}                                                              
\caption{For P72, partial radial distribution functions at low (a) and high (b) density
according to the time-averaged size (see text). 
In (b) we include the $g(r)$ of the M72 system.
Snapshots: in (a),                
green, red and blue correspond to small, middle and large SCNPs respectively;
in (b) only the small ones are shown (each in a different color).}
\label{grparz}
\end{figure}

The P30 system has a lower degree of order than the more tightly cross-linked P72 system.
This is not suprising since the higher deformability of the P30 SCNPs facilitates stronger fluctuations of the local correlations.
Interestingly and in a counterintuitive fashion, the polydisperse P72 system has a {\it higher} degree of order than
its monodisperse counterpart M72 at the same density. 
This is demonstrated by the higher peak in the $g(r)$ of P72 respect to M72.
To shed light on the origin of this unexpected feature, we analyze in more detail the correlations between the intrinsically polydisperse SCNPs in the P72 system. We divide the SCNPs in three `species' (`small', `middle', and `large'). 
The small and large SCNPs are defined as the $16\%$ of particles with, respectively, the lowest
and largest values  of the time-averaged diameter of gyration $\bar{D}_{\rm g0}$. The middle SCNPs are the remaining 68\%.
Fig.~\ref{grparz} shows results, for the P72 system, of the partial $g(r)$'s of the three species.
At low concentration (panel (a)) the three functions are very similar.
The inset shows a snapshot of the solution,
where the three species indeed seem to be uniformly distributed.
However, by increasing the concentration, the small particles start to aggregate and form clusters. This is highlighted 
in panel (b) (density $\rho = 1.57$, above the overlap concentration),
both by the snapshot (where only the small SCNPs are displayed), and by the sharp main peak of the partial $g(r)$
of the small particles. In this panel we also include the $g(r)$ of the monodisperse system M72
at the same  $\rho = 1.57$. Differences
between this and the $g(r)$ of the 68\% middle particles of the P72 system are negligible.
On the other hand, differences are much more pronounced with the $g(r)$ of the 16\% smallest particles than with the same fraction of the largest particles. 
Therefore we conclude that the higher degree of ordering
in the polydisperse P72 respect to the monodisperse M72 originates from the aggregation and clustering
of the small particles in the former.

\section{Dynamics}

\begin{figure}
\includegraphics[width=.91\linewidth]{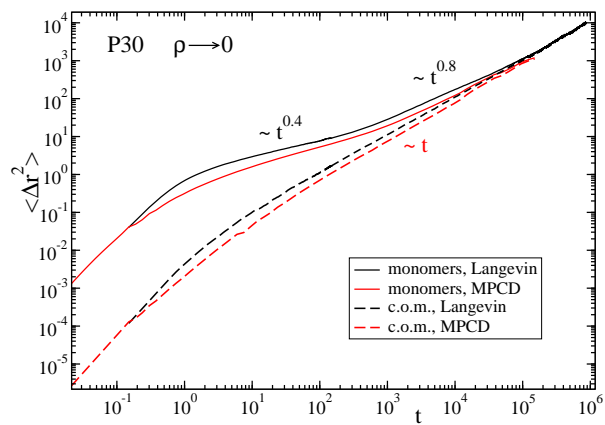}
\caption{Mean squared displacement of the monomers (solid curves) and centers-of-mass (c.o.m, dashed curves) for P30-SCNPs at infinite dilution. Data are shown for simulations with (MPCD, red curves) and without (Langevin dynamics, black curves) hydrodynamic interactions. Subdiffusive and diffusive power-laws are indicated for comparison.}
\label{msd-lang-mpcd}
\end{figure}

As mentioned in Section 2, hydrodynamic interactions have been neglected and all
the simulations  have been performed under LD, though for testing this approximation
MPCD simulations have also been performed for the P30 system at $\rho \rightarrow 0$.
Fig.~\ref{msd-lang-mpcd} shows the corresponding mean squared displacements ($\langle \Delta r^2(t) \rangle$, MSD), 
both for monomers and centers-of-mass, obtained 
by LD and MPCD simulations. 
The results confirm that hydrodynamic interactions play a minor role. 
Only small differences are found between LD and MPCD at infinite dilution. Therefore such differences will be 
negligible at the high concentrations where the slow dynamics arise, and the use of LD is justified.
A detailed characterization of the intramolecular dynamics is beyond the scope of this work and will be presented
elsewhere. Still, the results in Fig.~\ref{msd-lang-mpcd} for the MSD of the monomers, exhibiting
two apparent subdiffusive regimes  prior to the final transition to diffusion ($\langle\Delta r^2\rangle \propto t$),
anticipate a complex character of the intramolecular dynamics.

\begin{figure}
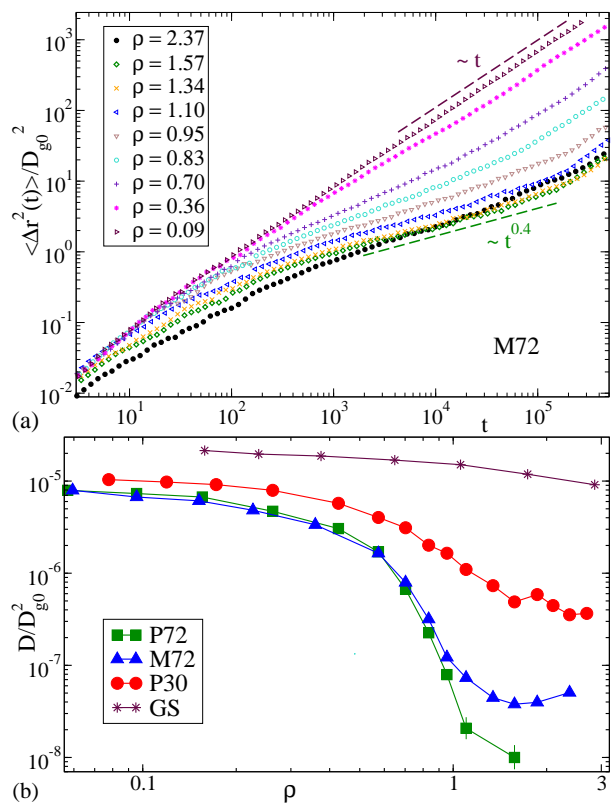

\begin{center}
\includegraphics[width=.92\linewidth]{Figure8a.eps}

\includegraphics[width=.92\linewidth]{Figure8b.eps}
\end{center}
\caption{(a): Normalised MSD of the centers-of-mass for the M72 solution at different densities.
Top and bottom dashed lines indicate diffusive and subdiffusive behaviour, respectively.
(b): Diffusivity vs. density for the three investigated systems of globular SCNPs.
For comparison we include results for sparse SCNPs synthesised in good solvent (GS).
Unless explicitly indicated, error bars are smaller than the symbol size.}
\label{msd}
\end{figure}

Now we investigate the dynamic consequences of the observed structural anomalies at concentrated solutions
In Fig.~\ref{msd}a we show the MSD of the SCNP centers-of-mass for the M72 solution, up to the highest density investigated.
The MSD data are normalised by $D_{\rm g0}^2 = \langle D_{\rm g0}^2 \rangle$ in order to show displacements in terms of the SCNP size. 
In standard colloidal fluids at high densities, particles can be mutually trapped by their neighbours over several time decades. 
This is the well-known caging effect that leads, after the short-time regime, to the appearance of a plateau
in the MSD versus time $t$, and whose duration increases by increasing concentration.
At longer times, particles escape from the cage and reach the diffusive regime.
However, within the investigated concentration range, no well-defined plateau is found in the solutions of SCNPs.
Instead, as a consequence of the deformable and penetrable character of the SCNPs,  
the MSD exhibits a soft caging regime, which eventually ends in the diffusive regime at long times.
The same qualitative behaviour is found for the three systems investigated.

\begin{figure}[t]
\includegraphics[width=0.99\linewidth]{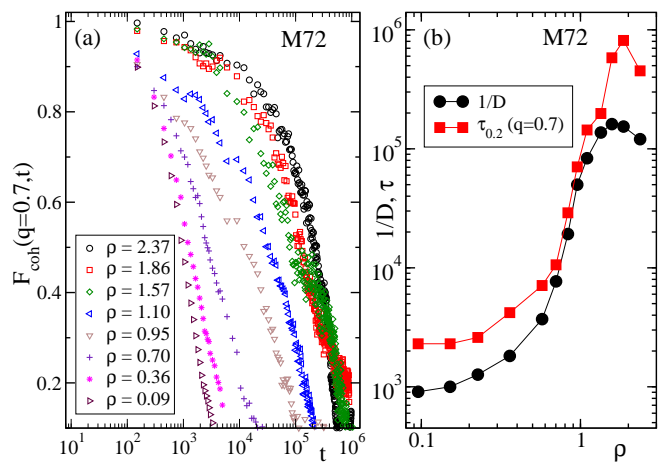}
\caption{(a): Density dependence of the coherent scattering function of the M72 system.
(b) Density dependence of the inverse diffusivities (circles) and the relaxation times (squares) 
of the coherent scattering functions for the M72 system. Error bars in (b) are smaller than the
symbol size.}
\label{taud}                                                                                            
\end{figure}

Dynamic differences between the three systems can be quantified by analysing the diffusivity $D$.
This is determined as the long-time limit of $\langle \Delta r^2(t)\rangle/6t$, for the densities at which the linear regime is reached within the simulation time.
Fig.~\ref{msd}b shows the density-dependence of the normalized diffusivity of the SCNP center-of-mass, $D/D^2_{\rm g0}$. 
At low densities $\rho \ll \rho^{\ast}$ collisions are relatively unfrequent and increasing concentration just leads to a weak reduction of the diffusivity.
While approaching the overlap density, we see a strong decrease of the mobility of the SCNPs.
Concomitant with the emergence of the soft caging regime in the MSD, the diffusivity exhibits the characteristic behaviour
of hard colloid solutions, i.e., a sharp drop in a narrow density interval close to the overlap concentration. 
Still, a steeper behaviour is found for the more tightly cross-linked systems (P72 and M72) than for the more deformable P30,
in close analogy with observations in microgels of tunable softness \cite{nature-microgels}.
For comparison,  we display in Fig.~\ref{msd}b the diffusivity for
solutions of SCNPs (25 \% of cross-linked groups) synthesised in good solvent.
These SCNPs (denoted as GS) show open sparse topologies \cite{moreno2013}.
As expected for the highly deformable GS-SCNPs, they are much more penetrable than the globular SCNPs and produce a much weaker excluded volume effect. Hence,  
in the same range of normalized density they show a much weaker
reduction of mobility than the globular SCNPs.

A striking behaviour is found by increasing the concentration of the globular SCNPs beyond the overlap density $\rho^{\ast} =1$. 
The dynamic counterpart of the loss of structural order highlighted by $g(r)$ is a much weaker density 
dependence of $D$ for $\rho > \rho^{\ast}$, and 
eventually a reentrant behaviour (see the diffusivity minima at $\rho \sim 1.6$ in Fig.~\ref{msd}b).   
Similar findings are obtained by analysing the normalised coherent scattering function of the SCNPs 
centers-of-mass, $F_{\rm coh}(q,t)$, which probes the slowing down of the collective motion. 
We compute $F_{\rm coh}(q,t)$ at the wavevector $q$ of the maximum of the 
static structure factor of the centers-of-mass, at the highest density investigated. 
Fig.~\ref{taud}a shows the results for the M72 system.
We calculate the relaxation time $\tau$ as $F_{\rm coh}(q,t=\tau)=0.2$
(we denote $\tau = \tau_{0.2}$). Fig.~\ref{taud}b  
shows results for the density dependence of $\tau_{0.2}$ in the M72 system. We include the corresponding values of
$1/D$ for comparison. As observed for the diffusivity, we find a similar reentrance in the relaxation times.

This dynamic reentrance by increasing concentration resembles that observed in models of ultrasoft particles \cite{PhysRevE.79.031203,PhysRevLett.90.238301,pamies-hertz,berthier-hertz}. 
To the best of our knowledge, the present 
results constitute the first observation of this dynamic anomaly in a real, monomer-resolved soft colloid with purely repulsive interactions.
In summary, the former observations open up the possibility of modifying and controlling the dynamical behaviour of the solutions by tuning the compactness and deformability
of the SCNPs through their degree of cross-linking. In particular, they allow for the realisation of dynamical anomalies proposed by models of ultrasoft particles. This result is highly non-trivial since such anomalies arise in our system {\it beyond} the overlap density. Indeed our monomer-resolved model accounts for intramolecular shrinkage and deformation at such densities,
a feature not considered by the mentioned models, which are based on density-independent interactions between structureless single-particles.

\begin{figure}
\includegraphics[width=0.97\linewidth]{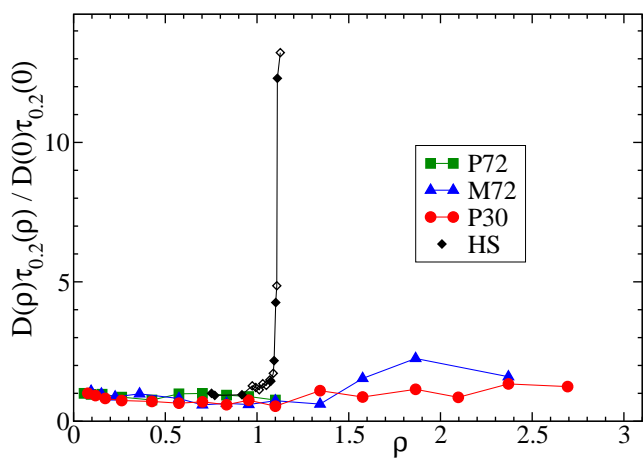}
\caption{Normalised product $D\tau$ vs. $\rho$ for 
P72, M72 and P30, and for the hard sphere (HS) system.
Data for HS are taken from Refs.~\cite{PhysRevLett.115.128302} (filled diamonds) and ~\cite{kumar-hs} (open diamonds).}
\label{stokein}                                                                                            
\end{figure}

So far we have disclosed the remarkable tunability of the dynamical behaviour of SCNPs by changing their softness/deformability. 
As outcome, the system is very promising in terms of clarifying important issues connected to the dynamics of soft colloids
with slow relaxation.
Inspired by recent work \cite{PhysRevLett.115.128302}
we test the validity of the Stokes-Einstein (SE) relation. 
A usual fingerprint of fluids showing slow structural relaxation
is the breakdown of the SE relation \cite{bonn-se,kumar-hs}, namely the product of the viscosity 
and the diffusivity is no longer constant as relaxation times increase by e.g.,
increasing density or decreasing temperature.                                                                       
This feature is attributed to the presence of dynamical heterogeneities,                        
which are weighted in a different fashion by the diffusivity and viscosity observables.
In Ref.~\cite{PhysRevLett.115.128302} this question was investigated 
in a system of polydisperse star-like micelles. 
Unlike in the archetype system of hard-sphere (HS) colloids 
(see Refs.~\cite{PhysRevLett.115.128302,kumar-hs} and data in Fig.~\ref{stokein}), 
no breakdown of the SE was observed up to the overlap density. 
This was assigned to the ultrasoft character of the interaction between the micelles \cite{PhysRevLett.115.128302},
as suggested by models of ultrasoft particles displaying weaker dynamic heterogeneity
than the hard counterparts \cite{PhysRevLett.106.015701}.   
In Fig.~\ref{stokein} we show the product of the relaxation time $\tau$ (proportional to the viscosity)
and the diffusivity, normalised to the value at infinite dilution.
Data in Fig.~\ref{stokein} demonstrate that, unlike in HS, the SE relation holds for the globular SCNPs,
in agreement with the results of Ref.~\cite{PhysRevLett.115.128302} for star-like soft colloids,
and extending them over a much broader density range,
far above the overlap concentration (unexplored up to now).

\section{Conclusions}
  
In summary, we have unveiled globular SCNPs as a very promising and versatile system.
Well-defined synthesis protocols can produce objects with different degrees of compactness.
By means of such realistic, experimentally accessible and flexible system,
it is possible to target different rheological properties, and in particular anomalous features
proposed by models of ultrasoft particles.
These include reentrant behavior in the density dependence of the structural and dynamical correlations
of the SCNPs, a soft caging regime, and weak dynamic heterogeneity far beyond the overlap density.
Due to its tunable soft colloidal character, covering the gap between polymers and hard colloids, globular SCNPs  
represent an optimal platform for a deep comprehension of the mechanisms behind diffusion of soft nano-objects in crowded environments.
\\
\\
\noindent{\large{\bf{Acknowledgements}}}
\normalsize
\\\\                                     
We acknowledge financial support from the projects MAT2015-63704-P and IT-654-13 (Spain). 


\begin{mcitethebibliography}{36}
\providecommand*{\natexlab}[1]{#1}
\providecommand*{\mciteSetBstSublistMode}[1]{}
\providecommand*{\mciteSetBstMaxWidthForm}[2]{}
\providecommand*{\mciteBstWouldAddEndPuncttrue}
  {\def\EndOfBibitem{\unskip.}}
\providecommand*{\mciteBstWouldAddEndPunctfalse}
  {\let\EndOfBibitem\relax}
\providecommand*{\mciteSetBstMidEndSepPunct}[3]{}
\providecommand*{\mciteSetBstSublistLabelBeginEnd}[3]{}
\providecommand*{\EndOfBibitem}{}
\mciteSetBstSublistMode{f}
\mciteSetBstMaxWidthForm{subitem}
{(\emph{\alph{mcitesubitemcount}})}
\mciteSetBstSublistLabelBeginEnd{\mcitemaxwidthsubitemform\space}
{\relax}{\relax}

\bibitem[Ouchi \emph{et~al.}(2011)Ouchi, Badi, Lutz, and Sawamoto]{cite-key}
M.~Ouchi, N.~Badi, J.-F. Lutz and M.~Sawamoto, \emph{Nat Chem}, 2011,
  \textbf{3}, 917--924\relax
\mciteBstWouldAddEndPuncttrue
\mciteSetBstMidEndSepPunct{\mcitedefaultmidpunct}
{\mcitedefaultendpunct}{\mcitedefaultseppunct}\relax
\EndOfBibitem
\bibitem[Altintas and Barner-Kowollik(2012)]{altintas2012single}
O.~Altintas and C.~Barner-Kowollik, \emph{Macromol. Rapid Commun.}, 2012,
  \textbf{33}, 958--971\relax
\mciteBstWouldAddEndPuncttrue
\mciteSetBstMidEndSepPunct{\mcitedefaultmidpunct}
{\mcitedefaultendpunct}{\mcitedefaultseppunct}\relax
\EndOfBibitem
\bibitem[Gonzalez-Burgos \emph{et~al.}(2015)Gonzalez-Burgos, Latorre-Sanchez,
  and Pomposo]{C5CS00209E}
M.~Gonzalez-Burgos, A.~Latorre-Sanchez and J.~A. Pomposo, \emph{Chem. Soc.
  Rev.}, 2015, \textbf{44}, 6122--6142\relax
\mciteBstWouldAddEndPuncttrue
\mciteSetBstMidEndSepPunct{\mcitedefaultmidpunct}
{\mcitedefaultendpunct}{\mcitedefaultseppunct}\relax
\EndOfBibitem
\bibitem[Mavila \emph{et~al.}(2016)Mavila, Eivgi, Berkovich, and
  Lemcoff]{doi:10.1021/acs.chemrev.5b00290}
S.~Mavila, O.~Eivgi, I.~Berkovich and N.~G. Lemcoff, \emph{Chem. Rev.}, 2016,
  \textbf{116}, 878--961\relax
\mciteBstWouldAddEndPuncttrue
\mciteSetBstMidEndSepPunct{\mcitedefaultmidpunct}
{\mcitedefaultendpunct}{\mcitedefaultseppunct}\relax
\EndOfBibitem
\bibitem[Hanlon \emph{et~al.}(2016)Hanlon, Lyon, and
  Berda]{doi:10.1021/acs.macromol.5b01456}
A.~M. Hanlon, C.~K. Lyon and E.~B. Berda, \emph{Macromolecules}, 2016,
  \textbf{49}, 2--14\relax
\mciteBstWouldAddEndPuncttrue
\mciteSetBstMidEndSepPunct{\mcitedefaultmidpunct}
{\mcitedefaultendpunct}{\mcitedefaultseppunct}\relax
\EndOfBibitem
\bibitem[Terashima \emph{et~al.}(2011)Terashima, Mes, De~Greef, Gillissen,
  Besenius, Palmans, and Meijer]{terashima2011}
T.~Terashima, T.~Mes, T.~F.~A. De~Greef, M.~A.~J. Gillissen, P.~Besenius,
  A.~R.~A. Palmans and E.~W. Meijer, \emph{J. Am. Chem. Soc.}, 2011,
  \textbf{133}, 4742--4745\relax
\mciteBstWouldAddEndPuncttrue
\mciteSetBstMidEndSepPunct{\mcitedefaultmidpunct}
{\mcitedefaultendpunct}{\mcitedefaultseppunct}\relax
\EndOfBibitem
\bibitem[Gillissen \emph{et~al.}(2012)Gillissen, Voets, Meijer, and
  Palmans]{gillissen2012single}
M.~A.~J. Gillissen, I.~K. Voets, E.~W. Meijer and A.~R.~A. Palmans,
  \emph{Polym. Chem.}, 2012, \textbf{3}, 3166--3174\relax
\mciteBstWouldAddEndPuncttrue
\mciteSetBstMidEndSepPunct{\mcitedefaultmidpunct}
{\mcitedefaultendpunct}{\mcitedefaultseppunct}\relax
\EndOfBibitem
\bibitem[Perez-Baena \emph{et~al.}(2013)Perez-Baena, Barroso-Bujans, Gasser,
  Arbe, Moreno, Colmenero, and Pomposo]{perez2013endowing}
I.~Perez-Baena, F.~Barroso-Bujans, U.~Gasser, A.~Arbe, A.~J. Moreno,
  J.~Colmenero and J.~A. Pomposo, \emph{ACS Macro Lett.}, 2013, \textbf{2},
  775--779\relax
\mciteBstWouldAddEndPuncttrue
\mciteSetBstMidEndSepPunct{\mcitedefaultmidpunct}
{\mcitedefaultendpunct}{\mcitedefaultseppunct}\relax
\EndOfBibitem
\bibitem[Sanchez-Sanchez \emph{et~al.}(2013)Sanchez-Sanchez, Akbari, Moreno,
  Lo~Verso, Arbe, Colmenero, and Pomposo]{sanchez2013design}
A.~Sanchez-Sanchez, S.~Akbari, A.~J. Moreno, F.~Lo~Verso, A.~Arbe, J.~Colmenero
  and J.~A. Pomposo, \emph{Macromol. Rapid Commun.}, 2013, \textbf{34},
  1681--1686\relax
\mciteBstWouldAddEndPuncttrue
\mciteSetBstMidEndSepPunct{\mcitedefaultmidpunct}
{\mcitedefaultendpunct}{\mcitedefaultseppunct}\relax
\EndOfBibitem
\bibitem[Altintas and Barner-Kowollik(2016)]{MARC:MARC201500547}
O.~Altintas and C.~Barner-Kowollik, \emph{Macromol. Rapid Commun.}, 2016,
  \textbf{37}, 29--46\relax
\mciteBstWouldAddEndPuncttrue
\mciteSetBstMidEndSepPunct{\mcitedefaultmidpunct}
{\mcitedefaultendpunct}{\mcitedefaultseppunct}\relax
\EndOfBibitem
\bibitem[Moreno \emph{et~al.}(2013)Moreno, Lo~Verso, Sanchez-Sanchez, Arbe,
  Colmenero, and Pomposo]{moreno2013}
A.~J. Moreno, F.~Lo~Verso, A.~Sanchez-Sanchez, A.~Arbe, J.~Colmenero and J.~A.
  Pomposo, \emph{Macromolecules}, 2013, \textbf{46}, 9748--9759\relax
\mciteBstWouldAddEndPuncttrue
\mciteSetBstMidEndSepPunct{\mcitedefaultmidpunct}
{\mcitedefaultendpunct}{\mcitedefaultseppunct}\relax
\EndOfBibitem
\bibitem[Pomposo \emph{et~al.}(2014)Pomposo, P\'erez-Baena, Lo~Verso, Moreno,
  Arbe, and Colmenero]{pomposo2014ACS}
J.~A. Pomposo, I.~P\'erez-Baena, F.~Lo~Verso, A.~J. Moreno, A.~Arbe and
  J.~Colmenero, \emph{ACS Macro Lett.}, 2014, \textbf{3}, 767--772\relax
\mciteBstWouldAddEndPuncttrue
\mciteSetBstMidEndSepPunct{\mcitedefaultmidpunct}
{\mcitedefaultendpunct}{\mcitedefaultseppunct}\relax
\EndOfBibitem
\bibitem[Lo~Verso \emph{et~al.}(2015)Lo~Verso, Pomposo, Colmenero, and
  Moreno]{C4SM02475C}
F.~Lo~Verso, J.~A. Pomposo, J.~Colmenero and A.~J. Moreno, \emph{Soft Matter},
  2015, \textbf{11}, 1369--1375\relax
\mciteBstWouldAddEndPuncttrue
\mciteSetBstMidEndSepPunct{\mcitedefaultmidpunct}
{\mcitedefaultendpunct}{\mcitedefaultseppunct}\relax
\EndOfBibitem
\bibitem[Merrifield(1963)]{doi:10.1021/ja00897a025}
R.~B. Merrifield, \emph{J. Am. Chem. Soc.}, 1963, \textbf{85}, 2149--2154\relax
\mciteBstWouldAddEndPuncttrue
\mciteSetBstMidEndSepPunct{\mcitedefaultmidpunct}
{\mcitedefaultendpunct}{\mcitedefaultseppunct}\relax
\EndOfBibitem
\bibitem[Basasoro \emph{et~al.}(2016)Basasoro, Gonzalez-Burgos, Moreno, Verso,
  Arbe, Colmenero, and Pomposo]{MARC:MARC201600139}
S.~Basasoro, M.~Gonzalez-Burgos, A.~J. Moreno, F.~L. Verso, A.~Arbe,
  J.~Colmenero and J.~A. Pomposo, \emph{Macromol. Rapid Commun.}, 2016,
  \textbf{37}, 1060--1065\relax
\mciteBstWouldAddEndPuncttrue
\mciteSetBstMidEndSepPunct{\mcitedefaultmidpunct}
{\mcitedefaultendpunct}{\mcitedefaultseppunct}\relax
\EndOfBibitem
\bibitem[Perez-Baena \emph{et~al.}(2014)Perez-Baena, Asenjo-Sanz, Arbe, Moreno,
  Verso, Colmenero, and Pomposo]{doi:10.1021/ma5017133}
I.~Perez-Baena, I.~Asenjo-Sanz, A.~Arbe, A.~J. Moreno, F.~L. Verso,
  J.~Colmenero and J.~A. Pomposo, \emph{Macromolecules}, 2014, \textbf{47},
  8270--8280\relax
\mciteBstWouldAddEndPuncttrue
\mciteSetBstMidEndSepPunct{\mcitedefaultmidpunct}
{\mcitedefaultendpunct}{\mcitedefaultseppunct}\relax
\EndOfBibitem
\bibitem[Kremer and Grest(1990)]{kremer1990dynamics}
K.~Kremer and G.~S. Grest, \emph{J. Chem. Phys.}, 1990, \textbf{92}, 5057\relax
\mciteBstWouldAddEndPuncttrue
\mciteSetBstMidEndSepPunct{\mcitedefaultmidpunct}
{\mcitedefaultendpunct}{\mcitedefaultseppunct}\relax
\EndOfBibitem
\bibitem[Berthier \emph{et~al.}(2010)Berthier, Moreno, and
  Szamel]{berthier-hertz}
L.~Berthier, A.~J. Moreno and G.~Szamel, \emph{Phys. Rev. E}, 2010,
  \textbf{82}, 060501\relax
\mciteBstWouldAddEndPuncttrue
\mciteSetBstMidEndSepPunct{\mcitedefaultmidpunct}
{\mcitedefaultendpunct}{\mcitedefaultseppunct}\relax
\EndOfBibitem
\bibitem[Blanco \emph{et~al.}(2015)Blanco, Shen, and Ferrari]{blanco}
E.~Blanco, H.~Shen and M.~Ferrari, \emph{Nat. Biotechnol.}, 2015, \textbf{33},
  941--951\relax
\mciteBstWouldAddEndPuncttrue
\mciteSetBstMidEndSepPunct{\mcitedefaultmidpunct}
{\mcitedefaultendpunct}{\mcitedefaultseppunct}\relax
\EndOfBibitem
\bibitem[Zhang \emph{et~al.}(2012)Zhang, Cao, Li, Ella-Menye, Bai, and
  Jiang]{zhang}
L.~Zhang, Z.~Cao, Y.~Li, J.-R. Ella-Menye, T.~Bai and S.~Jiang, \emph{ACS
  Nano}, 2012, \textbf{6}, 6681--6686\relax
\mciteBstWouldAddEndPuncttrue
\mciteSetBstMidEndSepPunct{\mcitedefaultmidpunct}
{\mcitedefaultendpunct}{\mcitedefaultseppunct}\relax
\EndOfBibitem
\bibitem[Vlassopoulos and Cloitre(2014)]{Vlassopoulos2014561}
D.~Vlassopoulos and M.~Cloitre, \emph{Curr. Opin. Colloid Interface Sci.},
  2014, \textbf{19}, 561 -- 574\relax
\mciteBstWouldAddEndPuncttrue
\mciteSetBstMidEndSepPunct{\mcitedefaultmidpunct}
{\mcitedefaultendpunct}{\mcitedefaultseppunct}\relax
\EndOfBibitem
\bibitem[Berendsen \emph{et~al.}(1995)Berendsen, van~der Spoel, and van
  Drunen]{BERENDSEN199543}
H.~Berendsen, D.~van~der Spoel and R.~van Drunen, \emph{Comput. Phys. Commun.},
  1995, \textbf{91}, 43 -- 56\relax
\mciteBstWouldAddEndPuncttrue
\mciteSetBstMidEndSepPunct{\mcitedefaultmidpunct}
{\mcitedefaultendpunct}{\mcitedefaultseppunct}\relax
\EndOfBibitem
\bibitem[Pooley \emph{et~al.}(2005)Pooley, , and
  Yeomans]{doi:10.1021/jp046040x}
C.~M. Pooley,  and J.~M. Yeomans, \emph{J. Phys. Chem. B}, 2005, \textbf{109},
  6505--6513\relax
\mciteBstWouldAddEndPuncttrue
\mciteSetBstMidEndSepPunct{\mcitedefaultmidpunct}
{\mcitedefaultendpunct}{\mcitedefaultseppunct}\relax
\EndOfBibitem
\bibitem[Mussawisade \emph{et~al.}(2005)Mussawisade, Ripoll, Winkler, and
  Gompper]{mussawisade-hydro}
K.~Mussawisade, M.~Ripoll, R.~G. Winkler and G.~Gompper, \emph{J. Chem. Phys.},
  2005, \textbf{123}, year\relax
\mciteBstWouldAddEndPuncttrue
\mciteSetBstMidEndSepPunct{\mcitedefaultmidpunct}
{\mcitedefaultendpunct}{\mcitedefaultseppunct}\relax
\EndOfBibitem
\bibitem[Singh \emph{et~al.}(2014)Singh, Huang, Westphal, Gompper, and
  Winkler]{sing-hydro}
S.~P. Singh, C.-C. Huang, E.~Westphal, G.~Gompper and R.~G. Winkler, \emph{J.
  Chem. Phys.}, 2014, \textbf{141}, year\relax
\mciteBstWouldAddEndPuncttrue
\mciteSetBstMidEndSepPunct{\mcitedefaultmidpunct}
{\mcitedefaultendpunct}{\mcitedefaultseppunct}\relax
\EndOfBibitem
\bibitem[Likos \emph{et~al.}(1998)Likos, L\"owen, Watzlawek, Abbas,
  Jucknischke, Allgaier, and Richter]{PhysRevLett.80.4450}
C.~N. Likos, H.~L\"owen, M.~Watzlawek, B.~Abbas, O.~Jucknischke, J.~Allgaier
  and D.~Richter, \emph{Phys. Rev. Lett.}, 1998, \textbf{80}, 4450--4453\relax
\mciteBstWouldAddEndPuncttrue
\mciteSetBstMidEndSepPunct{\mcitedefaultmidpunct}
{\mcitedefaultendpunct}{\mcitedefaultseppunct}\relax
\EndOfBibitem
\bibitem[Mohanty and Richtering(2008)]{doi:10.1021/jp808203d}
P.~S. Mohanty and W.~Richtering, \emph{J. Phys. Chem. B}, 2008, \textbf{112},
  14692--14697\relax
\mciteBstWouldAddEndPuncttrue
\mciteSetBstMidEndSepPunct{\mcitedefaultmidpunct}
{\mcitedefaultendpunct}{\mcitedefaultseppunct}\relax
\EndOfBibitem
\bibitem[Foffi \emph{et~al.}(2003)Foffi, Sciortino, Tartaglia, Zaccarelli,
  Lo~Verso, Reatto, Dawson, and Likos]{PhysRevLett.90.238301}
G.~Foffi, F.~Sciortino, P.~Tartaglia, E.~Zaccarelli, F.~Lo~Verso, L.~Reatto,
  K.~A. Dawson and C.~N. Likos, \emph{Phys. Rev. Lett.}, 2003, \textbf{90},
  238301\relax
\mciteBstWouldAddEndPuncttrue
\mciteSetBstMidEndSepPunct{\mcitedefaultmidpunct}
{\mcitedefaultendpunct}{\mcitedefaultseppunct}\relax
\EndOfBibitem
\bibitem[P\`{a}mies \emph{et~al.}(2009)P\`{a}mies, Cacciuto, and
  Frenkel]{pamies-hertz}
J.~C. P\`{a}mies, A.~Cacciuto and D.~Frenkel, \emph{J. Chem. Phys.}, 2009,
  \textbf{131}, \relax
\mciteBstWouldAddEndPuncttrue
\mciteSetBstMidEndSepPunct{\mcitedefaultmidpunct}
{\mcitedefaultendpunct}{\mcitedefaultseppunct}\relax
\EndOfBibitem
\bibitem[Jacquin and Berthier(2010)]{jacquin-harmon}
H.~Jacquin and L.~Berthier, \emph{Soft Matter}, 2010, \textbf{6},
  2970--2974\relax
\mciteBstWouldAddEndPuncttrue
\mciteSetBstMidEndSepPunct{\mcitedefaultmidpunct}
{\mcitedefaultendpunct}{\mcitedefaultseppunct}\relax
\EndOfBibitem
\bibitem[Mattsson \emph{et~al.}(2009)Mattsson, Wyss, Fernandez-Nieves,
  Miyazaki, Hu, Reichman, and Weitz]{nature-microgels}
J.~Mattsson, H.~M. Wyss, A.~Fernandez-Nieves, K.~Miyazaki, Z.~Hu, D.~R.
  Reichman and D.~A. Weitz, \emph{Nature}, 2009, \textbf{462}, 83--86\relax
\mciteBstWouldAddEndPuncttrue
\mciteSetBstMidEndSepPunct{\mcitedefaultmidpunct}
{\mcitedefaultendpunct}{\mcitedefaultseppunct}\relax
\EndOfBibitem
\bibitem[Krekelberg \emph{et~al.}(2009)Krekelberg, Kumar, Mittal, Errington,
  and Truskett]{PhysRevE.79.031203}
W.~P. Krekelberg, T.~Kumar, J.~Mittal, J.~R. Errington and T.~M. Truskett,
  \emph{Phys. Rev. E}, 2009, \textbf{79}, 031203\relax
\mciteBstWouldAddEndPuncttrue
\mciteSetBstMidEndSepPunct{\mcitedefaultmidpunct}
{\mcitedefaultendpunct}{\mcitedefaultseppunct}\relax
\EndOfBibitem
\bibitem[Gupta \emph{et~al.}(2015)Gupta, Stellbrink, Zaccarelli, Likos,
  Camargo, Holmqvist, Allgaier, Willner, and Richter]{PhysRevLett.115.128302}
S.~Gupta, J.~Stellbrink, E.~Zaccarelli, C.~N. Likos, M.~Camargo, P.~Holmqvist,
  J.~Allgaier, L.~Willner and D.~Richter, \emph{Phys. Rev. Lett.}, 2015,
  \textbf{115}, 128302\relax
\mciteBstWouldAddEndPuncttrue
\mciteSetBstMidEndSepPunct{\mcitedefaultmidpunct}
{\mcitedefaultendpunct}{\mcitedefaultseppunct}\relax
\EndOfBibitem
\bibitem[Kumar \emph{et~al.}(2006)Kumar, Szamel, and Douglas]{kumar-hs}
S.~K. Kumar, G.~Szamel and J.~F. Douglas, \emph{J. Chem. Phys.}, 2006,
  \textbf{124}, year\relax
\mciteBstWouldAddEndPuncttrue
\mciteSetBstMidEndSepPunct{\mcitedefaultmidpunct}
{\mcitedefaultendpunct}{\mcitedefaultseppunct}\relax
\EndOfBibitem
\bibitem[Bonn and Kegel(2003)]{bonn-se}
D.~Bonn and W.~K. Kegel, \emph{J. Chem. Phys.}, 2003, \textbf{118},
  2005--2009\relax
\mciteBstWouldAddEndPuncttrue
\mciteSetBstMidEndSepPunct{\mcitedefaultmidpunct}
{\mcitedefaultendpunct}{\mcitedefaultseppunct}\relax
\EndOfBibitem
\bibitem[Ikeda and Miyazaki(2011)]{PhysRevLett.106.015701}
A.~Ikeda and K.~Miyazaki, \emph{Phys. Rev. Lett.}, 2011, \textbf{106},
  015701\relax
\mciteBstWouldAddEndPuncttrue
\mciteSetBstMidEndSepPunct{\mcitedefaultmidpunct}
{\mcitedefaultendpunct}{\mcitedefaultseppunct}\relax
\EndOfBibitem
\end{mcitethebibliography}

\providecommand*{\mcitethebibliography}{\thebibliography}
\csname @ifundefined\endcsname{endmcitethebibliography}
{\let\endmcitethebibliography\endthebibliography}{}

\end{document}